\documentclass[12pt,a4paper]{article}
\usepackage{cmap}
\usepackage[T2A]{fontenc}
\usepackage[cp1251]{inputenc}
\usepackage[english, russian]{babel}
\usepackage{amsfonts,amsmath,amssymb,amsthm,longtable,hhline}
\usepackage{bm}
\usepackage{cite}
\usepackage{color,soul}
\usepackage{color,xcolor}
\usepackage{multicol}
\usepackage{tabularx,caption,multirow}
\usepackage{ulem}
\usepackage{titlesec}

\makeatletter
\renewcommand{\@biblabel}[1]{#1.} 
\makeatother
\makeatletter

\newcommand{\clh}[1]{\colorbox{yellow}{#1}}%
\newcommand{\clhp}[1]{\colorbox{yellow}{\parbox{\textwidth}{#1}}}%
\newlength\aptextwidth
\definecolor{BrickRed}{rgb}{0.588,0.098,0.055}

\def\noblue#1{\ifmmode \text{#1}\else #1\fi}
\def\noclh#1{\ifmmode \text{#1}\else #1\fi}
\def\rem#1{}

\let\ge=\geqslant

\def\ttable#1. #2{\begin{table}[t]\tablehat{#1}{#2}}
\def\mtable#1. #2{\begin{table}[hbtp]\tablehat{#1}{#2}}
\def\ptable#1. #2{\begin{table}[p]\tablehat{#1}{#2}}
\def\tablehat#1#2{\centering\small \vbox{\parindent=0pt
  \leftskip=0pt plus.5\hsize \rightskip=\leftskip \parfillskip=0pt
  ТАБЛИЦА #1\\ #2}\nobreak\medskip\medskip }
\def\texendtable{\end{table}}
\def\notation{\par\ifnum\lastpenalty<25000 \bigbreak \fi
  \noindent\triangle\enspace\ignorespaces}

\def\hline{\noalign{\hrule}}

\def\eqnitemskip{\ifhmode \else \par
  \ifnum\lastpenalty>24999
    \ifnum\lastpenalty=25004 \fi
  \else \medbreak \fi \fi }
\def\eqnitem #1. {\eqnitemskip
  {\setbox0=\hbox{$#1^\circ$.\enspace}%
  \ifdim\wd0>\parindent \box0\ignorespaces \else
  \hbox to\parindent{\unhbox0\hss}\ignorespaces\fi}}
\def\eqnitemnobreak #1. {\noindent
  {\setbox0=\hbox{$#1^\circ$.\enspace}%
  \ifdim\wd0>\parindent \box0\ignorespaces \else
  \hbox to\parindent{\unhbox0\hss}\ignorespaces\fi}}
\newdimen\eqnparindent
\eqnparindent=\parindent
\def\eqnitem #1. {\eqnitemskip\noindent\hskip\eqnparindent $#1^\circ$.\enspace\ignorespaces }
\def\eqnitemnobreak #1. {\noindent\hskip\eqnparindent $#1^\circ$.\enspace\ignorespaces }
\def\simpleitem #1. {\eqnitemskip\noindent\hskip\eqnparindent #1.\enspace\ignorespaces }

\def\eqalignno#1{\displ@y \tabskip\centering
  \halign to\displaywidth{\hfil$\@lign\displaystyle{##}$\tabskip\z@skip
    &$\@lign\displaystyle{{}##}$\hfil\tabskip\centering
    &\llap{$\@lign\eqnofont##$}\tabskip\z@skip\crcr
    #1\crcr}}
\let\eqalignno=\eqalignm
\def\eqcenter#1{\displ@y \tabskip\centering
  \halign{\hfil$\displaystyle{##}$\hfil\crcr
    #1\crcr}}
\def\eqcenterno#1{\displ@y \tabskip\centering
  \halign to\displaywidth{\hfil$\@lign\displaystyle{##}$\hfil
    \tabskip\centering&\llap{$\@lign\eqnofont##$}\tabskip\z@skip\crcr
    #1\crcr}}
\def\texcases#1{\left\{\,\vcenter{\normalbaselines\m@th
    \ialign{$##\hfil$&\quad##\hfil\crcr#1\crcr}}\right.}
\def\Displaylines#1{\vcenter{\displ@y \tabskip\z@skip
  \halign{\hbox to\displaywidth{$\@lign\hfil\displaystyle##\hfil$}\crcr
    #1\crcr}}}

\def\tan{\mathop{\operator@font tg}\nolimits}

\makeatother

\titlespacing\section{0pt}{12pt plus 4pt minus 2pt}{0pt plus 2pt minus 2pt}
\titlespacing\subsection{0pt}{12pt plus 4pt minus 2pt}{0pt plus 2pt minus 2pt}

\begin{document}
\large 

\bigskip
\centerline{\bf\Large Exact solutions and reductions of}
\centerline{\bf\Large nonlinear Schr\"odinger equations with delay\clh{$^{*}$}}

\bigskip
\medskip

\centerline{Andrei D. Polyanin$^*$, Nikolay A. Kudryashov$^{**}$}
 \medskip
\centerline{$^*$~Ishlinsky Institute for Problems in Mechanics, Russian Academy of Sciences,}
\centerline{pr. Vernadskogo 101, bldg. 1, Moscow, 119526 Russia}
\smallskip
\centerline{$^{**}$~Department of Applied Mathematics, National Research Nuclear}
\centerline{University MEPhI, 31 Kashirskoe Shosse,  115409 Moscow, Russia}
\smallskip
\centerline{e-mails: polyanin@ipmnet.ru, nakudryashov@mephi.ru}
\bigskip

\let\thefootnote\relax\footnotetext{
\hskip-20pt\clhp{$^*$ This is a preprint of an article that will be published in the journal \textbf{Journal of Computational and Applied Mathematics, 2025, Vol.~462, 116477};
doi:\! 10.1016/j.cam.2024.116477.}}

\vspace{5.0ex}

For the first time, Schr\"odinger equations with cubic and more complex nonlinearities containing the unknown function with constant delay are analyzed.
The physical considerations that can lead to the appearance of a delay in such nonlinear equations and mathematical models are expressed.
One-dimensional non-symmetry reductions are described, which lead the studied partial differential equations with delay to simpler ordinary
differential equations and ordinary differential equations with delay.
New exact solutions of the nonlinear Schr\"odinger equation of the general form with delay, which are expressed in quadratures, are found.
To construct exact solutions, a combination of methods of generalized separation of variables and the method of functional constraints are used.
Special attention is paid to three equations with cubic nonlinearity, which allow simple solutions in elementary functions, as well as more complex exact solutions with generalized separation of variables.
Solutions representing a nonlinear superposition of two traveling waves, the amplitude of which varies periodically in time and space, are constructed.
Some more complex nonlinear Schr\"odinger equations of a general form with variable delay are also studied.
The results of this work can be useful for the development and improvement of mathematical models described by nonlinear Schr\"odinger equations with delay and related functional PDEs, and the obtained exact solutions can be used as test problems intended to assess the accuracy of numerical methods for integrating nonlinear equations of mathematical physics with delay.
\medskip

\textsl{Keywords\/}:
nonlinear Schr\"odinger equations,
PDEs with delay,
exact solutions,
solutions in quadratures,
solutions in elementary functions,
generalized separable solutions

\section{Introduction}

\subsection{Nonlinear Schr\"odinger equations and related PDEs}

It is well known that the nonlinear Schr\"odinger equation is one of popular nonlinear partial equation which is used in the many areas of theoretical physics including nonlinear optics, superconductivity and plasma physics. It takes the form
\cite{Ag1,Ag2,Ag3,Ag4,Ag5,Ag6,liu2019,fib2019}:
\begin{equation}
i\,u_t+k\,u_{xx}+f(|u|)u=0,
\label{Schrodinger-eq}
\end{equation}
where $u=u(x,t)$ is the desired complex-valued function of real variables, the quadrate of the module of which corresponds to the intensity of light, $t$ is the time, $x$ is the spatial variable, $f(|u|)$ is the potential, $k$ is a parameter of equation, $i^2=-1$.

The classical nonlinear Schr\"odinger equation with cubic nonlinearity, which is determined by the function
\begin{equation}
f(|u|)=b|u|^2,
\label{seq2}
\end{equation}
is well known in science.
Eq.~\eqref{Schrodinger-eq} describes mathematical models for wave propagation in essentially all sections of physics where wave processes are considered. However, this equation became especially popular after the theoretical and experimental substantiation of the application of the nonlinear Schr\"odinger equation in nonlinear optics \cite{UFN,Has1,Has2,Has3}.
When describing the propagation of pulses in an optical fiber, the expression with the second derivative is responsible for the dispersion of the pulse, the function $f(|u|)$ characterizes the interaction of the light pulse with the fiber material and determines the nonlinear dependence of the refractive index of light. For the classical nonlinear Schr\"odinger equation, function \eqref{seq2} corresponds to the quadratic dependence of the refractive index and is called Kerr nonlinearity.
The uniqueness of equation \eqref{Schrodinger-eq}--\eqref{seq2} is explained not
only by the fact that this equation is the basic equation for describing the processes of information transmission in optical medium, but also by the fact that it belongs to the class of integrable partial differential equations. The equation has an infinite number of conservation laws, B\"acklund transformations, and passes the Painlev\'e test \cite{Pain_1, Pain_2, Pain_3, Pain_4, Pain_5}. Exact solutions of the classical nonlinear Schr\"odinger equation and related equations of mathematical physics can be found, for example, in reference books \cite{polzai2012,kha2019,pol2025}. The Cauchy problem for equation \eqref{Schrodinger-eq}--\eqref{seq2} with an initial condition of a general form is solved by the method of the inverse scattering problem
\cite{Ag4, Ag5}.

Note that some exact solutions of nonlinear Schr\"odinger equation \eqref{Schrodinger-eq} for arbitrary functions $f(|u|)$ are given
in \cite{polzai2012,pol2025}. The main directions of generalizations of the nonlinear Schr\"odinger equation in describing the propagation of optical pulses in a nonlinear medium are aimed at expanding mathematical models taking into account high-order derivatives and additional nonlinear expres\-sions. The introduction of high-order derivatives for generalizations of the nonlinear Schr\"odinger equation is dictated by the need to take into account the high-order dispersion during pulse propagation. At the same time, taking into account the non-Kerr nonlinearity in the propagation of optical pulses makes it possible to take into account more complex processes in the propagation of optical solitons, which lead to more complex mathematical models, the analytical study of which leads to additional difficulties.
Solutions to related modified and more complex nonlinear Schr\"odinger-type equations one also can find in \cite{polzai2012,kha2019,pol2025,cho2011,Kudr_1, Kudr_2, Kudr_3, Kudr_4, Kudr_5, Kudr_6, Bis_1, Bis_2, Bis_3, Bis_4, Bis_5, Bis_6,azz2015,wan2020,lie2020,huy2021,wan2023,mur2025,pen2025,kud2023x,arn2023a,arn2023b}.

\subsection{Differential equations with delay}{\label{Delay_1}}

For mathematical modeling of many phenomena and processes exhibiting the properties of heredity (or aftereffect), when the rate of change of the desired value depends not only on its current value, but also on some value (or several values) in the past, differential equations with a delay are used. In biology and biomechanics, lag is associated with a limited rate of transmission of nerve and muscle reactions in living tissues. In medicine, in tasks on the spread of infectious diseases, the time delay is determined by the incubation period (the interval from the moment of infection to the appearance of the first signs of the disease). In the dynamics of populations, the delay is due to the fact that individuals
do not participate in reproduction immediately, but only after reaching reproductive age. In the dynamics of populations, the delay is due to the fact that individuals do not participate in reproduction immediately, but only after reaching reproductive age. In control theory, delays occur due to limited signal propagation speeds and limited speeds of technological processes. The most common partial
differential equations with a delay, methods for solving them and some applications
are described, for example, in books \cite{wu1996,polsorzhu2024}.

To formulate the simplest problems with aftereffect, ordinary differential equations (ODEs) are used, depending on the time $t$, which, in addition to the desired function $u(t)$, also contain the function $\bar u = u(t-\tau)$, where $\tau>0$ is the constant time delay. Partial differential equations of the reaction-diffusion type are often used to describe related, more complex, spatially inhomogeneous problems with delay
\cite{wu1996,polsorzhu2024}:
\begin{equation}
u_t=ku_{xx}+F(u,\bar u),\quad \bar u=u(x,t-\tau),
\label{reaction-eq}
\end{equation}
where $k > 0$ is the diffusion coefficient, $t$ is the time, $x$ is the spatial variable, $F(u,\bar u)$ is the kinetic function, and $\tau>0$ is the time delay
(further, $\tau$ is considered a constant unless otherwise specified).
The equation with delay \eqref{reaction-eq} is a natural generalization of the usual nonlinear reaction-diffusion equation
without delay with the function $F(u,\bar u)=f(u)$.
A special case of Eq. \eqref{reaction-eq} with $F(u,\bar u)=f(\bar u)$ admits simple physical interpretation: the transfer of matter in a locally nonequilibrium medium exhibits inertial properties, i.e. the system does not react to the impact instantly, as in the classical
locally equilibrium case, but with a time delay~$\tau$.


Many exact solutions of reaction-diffusion equations of the form \eqref{reaction-eq} with constant delay and related nonlinear parabolic equations with delay can be found in \cite{pol2025,polsorzhu2024,mel2008,pol7,polzhu2014,polsor2021b,aib2021}.
In \cite{pol2025,polsorzhu2024,polsor2021,polsor2023} some exact solutions of more complicated reaction-diffusion equations with variable delay of various types are described.
Exact solutions of nonlinear wave equations with constant and variable delay, which are formally obtained from \eqref{reaction-eq} by replacing the first time derivative $u_t$ with the second derivative $u_{tt}$, are presented in \cite{pol2025,polsorzhu2024,polzhu2014c,long2016}.

Similar reasoning can be extended to mathematical models described by nonlinear Schr\"odinger equations.
Although the speed of propagation of an electro\-magnetic wave through an optical fiber has a huge speed, the reaction of the optical fiber material has some inertia, which can lead to a delay. This inertia is especially evident in the propagation of ultrashort optical solitons for femtosecond pulses of less than 1ps.
In addition, as noted in \cite{Ag1}, taking into account forced Raman scattering when describing ultrashort pulses in an optical fiber led to the discovery of a new phenomenon called soliton frequency self-shift, which is directly related to the inertia of scattering and was explained by its
occurrence. It is established that this phenomenon generates a continuous shift in the carrier frequency of the optical soliton, in which its spectrum becomes so wide that the high-frequency components begin to transfer their energy to the low-frequency components.
The above leads to the expediency of taking into account the delay in the expressions for the potential in various generalizations and further modifications of the nonlinear Schr\"odinger equations.

\subsection{Terminology: what we mean when we say about exact solutions of nonlinear PDEs with delay} {\label{Exact}}

In this paper, exact solutions of nonlinear partial differential equations with delay are understood as the following solutions \cite{polsorzhu2024}:

(\textit{a})\enspace Solutions that are expressed in terms of elementary functions.

(\textit{b})\enspace Solutions that are expressed in quadratures, i.e. through elementary functions, functions included in the equation (this is necessary if the equation contains arbitrary or special functions) and indefinite
integrals.

(\textit{c})\enspace Solutions that are expressed through solutions of ordinary differential equations or systems of such equations.

(\textit{d})\enspace Solutions that are expressed in terms of solutions of ordinary differential equations with delay
or systems of such equations.

Various combinations of solutions described in (\textit{a})--(\textit{d}) are also allowed. In cases (\textit{a}) and (\textit{b}), the exact solution can be presented in explicit, implicit, or parametric form. It is important to note that the presence of a delay in the equations of mathematical physics significantly complicates the analysis of such equations and corresponding initial boundary value problems \cite{wu1996,polsorzhu2024}. In particular, PDEs with constant delay do not allow self-similar solutions \cite{polsorzhu2024}, which often have simpler PDEs without delay.

Note that exact solutions are mathematical standards that are often used as test problems to check the adequacy and assess the accuracy of numerical methods for integrating nonlinear PDEs and PDEs with delay. The most preferable for these purposes are simple solutions from Items (\textit{a}) and (\textit{b}). It is these exact solutions that are described further in this article.
\medskip

\textit{Remark 1.}
Of great interest are also solutions of nonlinear PDEs with delay, which are expressed through solutions of linear PDEs without delay (examples of such nonlinear PDEs with delay can be found in \cite{polsorzhu2024,pol2025,polsor2022}).
According to the terminology introduced in \cite{bro2023}, such equations can be called conditionally integrable nonlinear PDEs (perhaps more accurately, they can be called partially linearizable PDEs).
\medskip

It is important to note that for complex nonlinear PDEs depending on one or more arbitrary functions (and it is precisely such nonlinear PDE that is considered in this article), the vast majority of existing analytical methods for constructing exact solutions are either not applicable at all or are weakly effective.
Statistical processing of reference data \cite{polzai2012,pol2025} showed that at present the majority of exact solutions of such PDEs without delay were obtained by methods of generalized and functional separation of variables. Significantly fewer exact solutions of such equations were obtained by nonclassical methods of symmetry reductions and the method of differential constraints, which are much more difficult to use in practice.
In general, very few nonlinear PDEs without delay depending on arbitrary functions that admit exact closed-form solutions are known today.
For nonlinear PDEs with delay that depend on arbitrary functions, the most effective method for constructing exact solutions is the method of functional constraints \cite{polsorzhu2024,pol7}.

In this article, we will consider a nonlinear Schr\"odinger equation with delay that is much more complex than the nonlinear Schr\"odinger equation without delay \eqref{Schrodinger-eq}. In this equation, instead of the potential $f(u)$, we include a potential of the general form $F(|u|,|\bar u|)$, where $F$ is an arbitrary function of two arguments and $\bar u=u(x,t-\tau)$ is an unknown function with delay.
The presence of an arbitrary function of two arguments in the equation under consideration, as well as the delay, are highly complicating factors,
since for such equations the vast majority of existing analytical methods
for constructing exact solutions either do not work at all or do not work effectively enough.

\section{Nonlinear Schr\"odinger equation with delay. Special cases and transformations} {\label{NSE_delay}}

\subsection{Nonlinear Schr\"odinger equation of general form with delay}

Let us consider the one-dimensional nonlinear Schr\"odinger equation of the general form with delay
\begin{equation}
iu_t+ku_{xx}+F(|u|,|\bar u|)u=0,\quad \ \bar u=u(x,t-\tau),
\label{eq01}
\end{equation}
where $u=u(x,t)$ is the desired complex-valued function of real variables,
$k>0$ is a parameter, $F(z_1,z_2)$ is an arbitrary real continuous function of two variables, $\tau>0$ is the time delay, and $i^2=-1$. The nonlinear Schr\"odinger equation with delay \eqref{eq01} is generalization of the usual nonlinear Schr\"odinger equation without delay  \eqref{Schrodinger-eq}, which is determined by formula $F(|u|,|\bar u|)=f(|u|)$.

We will specifically highlight three functions
\begin{equation}
F(|u|,|\bar u|)=b|\bar u|^2,\quad F(|u|,|\bar u|)=b|u||\bar u|,\quad F(|u|,|\bar u|)=b_1|u|^2+b_2|\bar u|^2,
\label{eq01aa}
\end{equation}
that determine the potentials of equations of the form \eqref{eq01} with cubic nonlinearity, which in the absence of delay (i.e., at $\tau=0$) lead to the classical nonlinear Schr\"odinger equation \eqref{Schrodinger-eq}--\eqref{seq2}.
The nonlinear Schr\"odinger equations with delay with
quadratic potentials $F(|u|,|\bar u|)=b_1|u|^2+b_2|u||\bar u|+b_3|\bar u|^2$ will also be considered in Section \ref{s:5}.


\subsection{Transformations of the nonlinear Schr\"odinger equation with delay}

Let us consider the transformations that will be used further to analyze the nonlinear Schr\"odinger equation with delay.

$1^\circ$.
The transformation
\begin{equation}
x=C_1X+C_2,\quad t=C_3T+C_4,\quad u=U(X,T)\exp[i(C_5T+C_6)],
\label{eq01*}
\end{equation}
where $C_1,\dots,C_6$ are arbitrary real constants ($C_1,C_3\not=0$),
leads Eq. \eqref{eq01} to an equation of the similar form
\begin{equation}
\begin{gathered}
iU_T+kC_3C_1^{-2}U_{XX}+[C_3F(|U|,|\bar U|)-C_5]U=0,\\
\bar U=U(X,T-\tau_*),\quad \ \tau_*=\tau/C_3.
\end{gathered}
\label{eq01**}
\end{equation}

$2^\circ$.
Let us represent the desired complex-valued function in exponential form
\begin{equation}
u=re^{i\varphi},\quad \ r=|u|,
\label{eq02}
\end{equation}
where $r=r(x,t)\ge 0$ and $\varphi=\varphi(x,t)$ are real functions.

Differentiating \eqref{eq02}, we find the derivatives:
\begin{equation}
\begin{aligned}
u_t&=(r_t+ir\varphi_t)e^{i\varphi},\\
u_x&=(r_x+ir\varphi_x)e^{i\varphi},\\
u_{xx}&=[r_{xx}-r\varphi_x^2+i(2r_x\varphi_x+r\varphi_{xx})]e^{i\varphi}.
\end{aligned}
\label{eq03}
\end{equation}
Substitute \eqref{eq03} into \eqref{eq01}, and then divide all terms by $e^{i\varphi}$.
Having further equated the real and imaginary parts of the obtained relation to zero, we arrive at the following system of two real PDEs with delay:
\begin{equation}
\begin{aligned}
-r\varphi_t+kr_{xx}-kr\varphi_x^2+F(r,\bar r)r&=0,\quad \ \bar r=r(x,t-\tau),\\
r_t+2kr_x\varphi_x+kr\varphi_{xx}&=0.
\end{aligned}
\label{eq04}
\end{equation}

The system of functional PDEs \eqref{eq04} together with expression \eqref{eq02} will be used further to construct
exact solutions of the nonlinear Schr\"odinger equation with delay \eqref{eq01}.

\section{Exact solutions of the general nonlinear Schr\"odinger equation with delay}

Below we describe exact solutions of the nonlinear Schr\"odinger equation with delay \eqref{eq01} with a potential of the general form,
which is given by an arbitrary function of two variables $F(|u|,|\bar u|)$.
In order to construct exact solutions we use a combination of the methods of generalized separation of variables (see, for example, \cite{polzai2012,galsvi2007,polzhu2022}) and the method of functional constraints \cite{polsorzhu2024,pol7}.
Note that this paper uses exactly the same functional constraints as in work \cite{polkud2024}.
\medskip

\textit{Remark 2.}
To construct exact solutions of the nonlinear PDE with delay \eqref{eq01},
one can also use the principle of structural analogy of solutions, which is formulated as follows: exact solutions of simpler equations can serve as a basis for constructing solutions of more complex related equations (see, for example, \cite{polsorzhu2024,polsor2021}).
Namely, to construct exact solutions of equation with delay \eqref{eq01} one can use the structure of known exact solutions of the simpler related equation without delay \eqref{Schrodinger-eq} (these auxiliary exact solutions are given, for example, in \cite{polzai2012,pol2025}).
\medskip

Below we will first indicate the general structure of the solutions of the PDE system with delay \eqref{eq04}, and then present the main intermediate ODEs or delay ODEs and final formulas. All results are easily verified by direct substitution of the obtained exact solutions into the delay PDE \eqref{eq01} or system \eqref{eq04}.

\subsection{Traveling wave solutions with constant amplitude}

The system of equations \eqref{eq04} has the following simple exact solutions of the form
\begin{equation}
r=C_1,\quad \ \varphi=C_2x+C_3+Bt,\quad \ B=F(C_1,C_1)-kC_2^2,
\label{eq05ab}
\end{equation}
where $C_1$, $C_2$, $C_3$ are arbitrary real constants.
Substituting \eqref{eq05ab} into \eqref{eq02}, we obtain a traveling wave solution of the considered nonlinear PDE \eqref{eq01}:
$$
u=C_1e^{i(C_2x+C_3+Bt)},\quad \ B=F(C_1,C_1)-kC_2^2.
$$
This solution is periodic in space and time with a constant amplitude; it does not depend on the time delay $\tau$.

\subsection{Time-periodic solutions with amplitude depending on spatial variable}

The system of equations \eqref{eq04} admits more complex periodic in time $t$, but independent of the time delay $\tau$, exact solutions of the form
\begin{equation}
r=r(x),\quad \ \varphi=C_1t+\theta(x),
\label{eq05}
\end{equation}
where $C_1$ is an arbitrary constant, and the functions $r=r(x)$ and $\theta=\theta(x)$ are described by an ODE system of the form
\begin{equation}
\begin{aligned}
kr_{xx}''-kr(\theta_x')^2-C_1r+F(r,r)r&=0,\\
2r_x'\theta_x'+r\theta_{xx}''&=0.
\end{aligned}
\label{eq06}
\end{equation}
Integrating the second equation \eqref{eq06} twice, we consistently find
\begin{equation}
\theta'_x=C_2r^{-2},\quad \
\theta=C_2\int r^{-2}dx+C_3,
\label{eq07}
\end{equation}
where $C_2$ and $C_3$ are arbitrary constants.
Substituting \eqref{eq07} into the first equation \eqref{eq06}, we obtain a second-order nonlinear ODE of the autonomous form
\begin{equation}
\begin{aligned}
kr_{xx}''-kC_2^2r^{-3}-C_1r+F(r,r)r=0.
\end{aligned}
\label{eq08}
\end{equation}
The general solution of Eq. \eqref{eq08} can be represented in implicit form
\begin{equation}
\int\left[\frac {C_1}kr^2-C_2^2r^{-2}-\frac 2k\int rF(r,r)\,dr+C_4\right]^{-1/2}\,dr=C_5\pm x,
\label{eq08**}
\end{equation}
where $C_4$ and $C_5$ are arbitrary constants.

Thus, it is shown that the system of PDEs with delay \eqref{eq04} admits the exact solution \eqref{eq05}, which can be expressed in quadratures.

Note that for Schr\"odinger equations with cubic potentials, which are determined by the functions \eqref{eq01aa}, the left part \eqref{eq08**} can be expressed in terms of elliptic integrals.
\medskip

\textit{Remark 3.}
A more complex nonlinear Schr\"odinger equation with variable delay \eqref{eq01}, in which $\tau=\tau(x,t)>0$ is an arbitrary continuous function, also admits a solution of the form \eqref{eq02}, \eqref{eq05}, where the function $r=r(x)$ is described by ODE \eqref{eq08}, and the function $\theta=\theta(x)$ is found using the second relation~\eqref{eq07}.
Note that exact solutions of nonlinear reaction-diffusion equations with variable delay have been considered in \cite{pol2025,polsorzhu2024,polsor2021,polsor2023}.
\medskip

\subsection{Generalized separable solutions with amplitude depending on time}\label{ss:3.3}

Let us show that the system of PDEs \eqref{eq04} admits exact generalized separable solutions of the form
\begin{equation}
r=r(t),\quad \ \varphi=a(t)x^2+b(t)x+c(t).
\label{eq09}
\end{equation}
To do this we substitute \eqref{eq09} into \eqref{eq04}. As a result the first equation of the system of equations is reduced
to quadratic equation with respect to  $x$, the coefficients of which depend on time. By equating the functional coefficients of this quadratic equation to zero and adding the second equation of the system, which in
this case depends only on $t$, we obtain the following system of ODEs:
\begin{equation}
\begin{aligned}
a'_t&=-4ka^2,\\
b'_t&=-4kab,\\
c'_t&=-kb^2+F(r,\bar r),\\
r'_t&=-2kar.
\end{aligned}
\label{eq10}
\end{equation}
Here, the first three equations were divided by $r$ and the notation was introduced $\bar r=r(t-\tau)$.

First we integrate the first equation of system \eqref{eq10}, then the second and fourth, and finally the third.
As a  result we have
\begin{equation}
\begin{aligned}
r&=\frac{C_3}{\sqrt{t+C_1}},\quad a=\frac 1{4k(t+C_1)},\quad b=\frac {C_2}{2k(t+C_1)},\\
c&=\frac{C_2^2}{4k(t+C_1)}+\int F\left(\frac{C_3}{\sqrt{t+C_1}},\frac{C_3}{\sqrt{t-\tau+C_1}}\right)dt+C_4,
\end{aligned}
\label{eq11}
\end{equation}
where  $C_1$, $C_2$, $C_3$, and $C_4$ are arbitrary constants. Substituting the expressions \eqref{eq11} into~\eqref{eq09}, we obtain
\begin{equation}
r=\frac{C_3}{\sqrt{t+C_1}},\quad \varphi=\frac{(x+C_2)^2}{4k(t+C_1)}+\int F\left(\frac{C_3}{\sqrt{t+C_1}},\frac{C_3}{\sqrt{t-\tau+C_1}}\right)dt+C_4.
\label{eq12}
\end{equation}

Note that for the nonlinear Schr\"odinger equations with cubic potentials, which are defined by functions \eqref{eq01aa},
the integral on the right-hand side of the second expression \eqref{eq12} is expressed in term of elementary functions.
In particular, for the first function \eqref{eq01aa} the solutions \eqref{eq12} take the form
\begin{equation*}
r=\frac{C_3}{\sqrt{t+C_1}},\quad \varphi=\frac{(x+C_2)^2}{4k(t+C_1)}+bC_3^2\ln(t-\tau+C_1)+C_4.
\end{equation*}

\textit{Remark 4.}
A more complex nonlinear Schr\"odinger equation \eqref{eq01} with variable delay, where $\tau=\tau(t)>0$ is an arbitrary continuous function, also admits a solution of the form \eqref{eq02}, \eqref{eq05}, in which the functions $r=r(t)$ and $\theta=\theta(x,t)$ are found by formulas~\eqref{eq12} with $\tau=\tau(t)$.
For comparison, it is important to note that exact solutions (other than traveling wave solutions) of the general nonlinear reaction-diffusion equation with constant delay \eqref{reaction-eq} are currently unknown. Even more so are the solutions of this nonlinear reaction-diffusion equation with variable delay.

\subsection{Solutions that are nonlinear superpositions of traveling waves}

$1^\circ$.
The system of equations \eqref{eq04} admits exact solutions of the form
\begin{equation}
r=r(z),\quad \ \varphi=C_1t+C_2x+\theta(z),\quad \ z=x+\lambda t,
\label{eq05*}
\end{equation}
where  $C_1$, $C_2$, and $\lambda$ are arbitrary constants, which generalizes solution
\eqref{eq05}. The special case $C_1=C_2=0$ in \eqref{eq05*} defines the traveling wave solution.

Substituting \eqref{eq05*} into \eqref{eq04}, we obtain a mixed nonlinear system consisting of an ODE with delay
and an ODE without delay:
\begin{equation}
\begin{aligned}
-r(C_1+\lambda\theta'_z)+kr_{zz}''-kr(C_2+\theta'_z)^2+F(r,\bar r)r&=0,\quad \ \bar r=r(z-\lambda\tau),\\
\lambda r_z'+2kr_z'(C_2+\theta'_z)+kr\theta_{zz}''&=0.
\end{aligned}
\label{eq09*}
\end{equation}
The substitution $\xi=\theta'_z$ allows us to lower the order of this system by one.

\textit{Special case.}
In the particular case $\theta(z)=C_3$, $\lambda=-2kC_2$ for $C_2<0$ system \eqref{eq09*} is reduced to the single second-order ODE with constant delay
\begin{equation}
kr_{zz}''-(C_1+C_2^2k)r+F(r,\bar r)r=0,\quad \
\bar r=r(z-\tau_1),\quad \ \tau_1=-2kC_2\tau.
\label{eq09**}
\end{equation}

Note that solution \eqref{eq05*} with $\theta(z)=C_3$ can be interpreted as a nonlinear superposition of two traveling waves
(with different velocities in the variables $r$ and~$\varphi$).

$2^\circ$.
For the nonlinear Schr\"odinger equation with delay \eqref{eq01} with a potential of the special form
$$
F(|u|,|\bar u|)=f(|u|^2+|\bar u|^2),
$$
where $f(z)$ is an arbitrary function, in the ODE with delay \eqref{eq09**} one should set $F(r,\bar r)=f(r^2+\bar r^2)$.
In this case, Eq.~\eqref{eq09**} admits exact periodic solutions
\begin{equation}
r(z)=\beta_n|\sin(\sigma_n z+C_4)|,\quad \ n=0,\, 1,\, 2,\dots
\label{eq09a**}
\end{equation}
Here $C_4$ is an arbitrary constant, the parameters $\beta_n$ are found from the algebraic (transcendental) equation
$$
f(\beta_n^2)=k\sigma_n^2+C_1+C_2^2k,
$$
and the constants $\sigma_n$ are determined by the formulas
\begin{equation}
\sigma_n=\frac \pi {2\tau_1}(1+2n)=\frac \pi {2\lambda\tau}(1+2n),\quad \ n=0,\, 1,\, 2,\dots
\label{eq09b**}
\end{equation}

$3^\circ$.
The nonlinear Schr\"odinger equation with delay \eqref{eq01} with a cubic nonlinearity of the form
$$
F(|u|,|\bar u|)=b(|u|^2+|\bar u|^2),\quad \ b=\text{const},
$$
has an exact solution of the form \eqref{eq05*}, where $\theta(z)=C_3$, $\lambda=-2kC_2$ (for $C_2<0$),
the function $r(z)$ is given in \eqref{eq09a**}, and the constants $\beta_n$ and $\sigma_n$ are determined by the formulas
$$
\beta_n=\sqrt{\frac{k\sigma_n^2+C_1+C_2^2k}b},\quad \ \sigma_n=\frac \pi {2\tau_1}(1+2n)=\frac \pi {2\lambda\tau}(1+2n),\quad \ n=0,\, 1,\, 2,\dots
$$

\section{Generalized separable solutions of the Schr\"odinger equations with cubic nonlinearities and delay}

\subsection{Structure of exact solutions for the equations under consideration}

The nonlinear Schr\"odinger equations with delay \eqref{eq01} and cubic nonlinearities, which are defined by the functions \eqref{eq01aa}, admit
exact generalized separable solutions of the form
\begin{equation}
u(x,t)=(ax+c)\exp[i(\alpha x^2+\beta x+\gamma)],
\label{equ01}
\end{equation}
where the five defining functions $a=a(t)$, $c=c(t)$, $\alpha=\alpha(t)$, $\beta=\beta(t)$, and $\gamma=\gamma(t)$
are described by mixed systems of equations containing ODEs without delay and ODEs with delay.

Solution \eqref{equ01} in variables \eqref{eq02} is reduced to system \eqref{eq04}, in which one should put
\begin{equation}
r=ax+c,\quad \ \varphi=\alpha x^2+\beta x+\gamma.
\label{equ02}
\end{equation}
Substitute functions \eqref{equ02} into system \eqref{eq04}. Using the dependencies \eqref{eq01aa} and separating the variables in the resulting equations, we arrive at systems for defining functions. These systems for all three dependencies \eqref{eq01aa} are listed below.

\subsection{Mixed systems of equations for the defining functions}

$1^\circ$.
In the case $F(|u|,|\bar u|)=b|\bar u|^2$ the system of equations for the defining functions is written as follows:
\begin{equation}
\begin{aligned}
a'_t&=-6ka\alpha,\\
c'_t&=-2ka\beta-2kc\alpha,\\
\alpha'_t&=-4k\alpha^2+b\bar a^2,\\
\beta'_t&=-4k\alpha\beta+2b\bar a\bar c,\\
\gamma'_t&=-k\beta^2+b\bar c^2,
\end{aligned}
\label{equ03}
\end{equation}
where $\bar a=a(t-\tau)$ and $\bar c=c(t-\tau)$.

$2^\circ$.
For $F(|u|,|\bar u|)=bu|\bar u|$ the system of equations for the defining functions has the form
\begin{equation}
\begin{aligned}
a'_t&=-6ka\alpha,\\
c'_t&=-2ka\beta-2kc\alpha,\\
\alpha'_t&=-4k\alpha^2+ba\bar a,\\
\beta'_t&=-4k\alpha\beta+b(a\bar c+\bar a c),\\
\gamma'_t&=-k\beta^2+bc\bar c.
\end{aligned}
\label{equ04}
\end{equation}

$3^\circ$.
For $F(|u|,|\bar u|)=b_1|u|^2+b_2|\bar u|^2$ the system of equations for the defining functions is written as follows:
\begin{equation}
\begin{aligned}
a'_t&=-6ka\alpha,\\
c'_t&=-2ka\beta-2kc\alpha,\\
\alpha'_t&=-4k\alpha^2+b_1a^2+b_2\bar a^2,\\
\beta'_t&=-4k\alpha\beta+2b_1ac+2b_2\bar a\bar c,\\
\gamma'_t&=-k\beta^2+b_1c^2+b_2\bar c^2.
\end{aligned}
\label{equ05}
\end{equation}

The mixed systems \eqref{equ03}--\eqref{equ05} consisting of ordinary differential equations and ordinary differential equations with delay are significantly simpler than the considered nonlinear Schr\"odinger equations with delay \eqref{eq01}--\eqref{eq01aa}. These systems can, for example, be integrated by the numerical methods described in \cite{polsorzhu2024}.

In the special case $a=0$, the mixed systems \eqref{equ03}--\eqref{equ05} are completely integrated, since in this case solution \eqref{equ01} coincides, up to obvious re-notations, with solution \eqref{eq09} of the more general nonlinear Schr\"odinger equation with delay \eqref{eq01}.

Note that the solutions of nonlinear Schr\"odinger equations with delay \eqref{eq01} and cubic nonlinearities described in this section, which are determined by dependencies \eqref{eq01aa}, are generalized to the case of variable delay of the general form (in these equations and solutions, $\tau=\text{const}$ should
be replaced by $\tau=\tau(t)$).

\section{Solutions of Schr\"odinger equations with other nonlinearities and delay}\label{s:5}

\subsection{Weakly nonlinear Schr\"odinger equations with delay}

Let us first consider the weakly nonlinear Schr\"odinger equations with a delay and two potentials of a special but rather general form that
satisfy the condition
\begin{equation}
F(|u|,|u|)=\text{const}.
\label{equ06xyz}
\end{equation}
This condition means that at $\tau\to 0$, the nonlinear Schr\"odinger equation with
delay \eqref{eq01} degenerates into a linear Schr\"odinger equation without delay.
\medskip

$1^\circ$.
Using the system of equations \eqref{eq04} it is easy to verify that the nonlinear Schr\"odinger equation with delay \eqref{eq01} and a potential of the form
\begin{equation}
F(|u|,|\bar u|)=f(|u|-|\bar u|),
\label{equ06}
\end{equation}
where $f(z)$ is an arbitrary function, has solution \eqref{eq02} with defining functions linear in both independent variables
\begin{equation}
r=C_1x+2kC_1C_2t+C_3,\quad \ \varphi=-C_2x+[f(2kC_1C_2\tau)-kC_2^2]t+C_4,
\label{equ07}
\end{equation}
which include four arbitrary real constants $C_1$, $C_2$, $C_3$, and $C_4$.
Note that potential \eqref{equ06} satisfies condition \eqref{equ06xyz}.

Setting $f(z)=bz^2$ in \eqref{equ06}, we obtain
the nonlinear Schr\"odinger equation with delay \eqref{eq01} and cubic nonlinearity determined by the potential
$$
F(|u|,|\bar u|)=b(|u|-|\bar u|)^2.
$$

$2^\circ$.
The nonlinear Schr\"odinger equation with delay \eqref{eq01} and potential \eqref{equ06} has also other exact solutions of the form \eqref{eq02} with functions
\begin{equation}
r=C_1\sin(C_2x+\beta_n t+C_3), \quad \ \varphi=A_nx+B_nt+C_4,
\label{equ07x}
\end{equation}
where $C_1$, $C_2$, $C_3$, and $C_4$ are arbitrary real constants ($C_2\not=0$), and other parameters
are defined as follows
\begin{equation}
\beta_n = \frac{2\pi n}\tau, \quad A_n=-\frac{\beta_n}{2kC_2},\quad B_n=-kC_2^2-\frac{\beta_n^2}{4kC_2^2}+f(0),\quad n=0,\,\pm 1,\, \pm 2,\,\dots
\label{equ07y}
\end{equation}

The solutions which are described by formulas \eqref{eq02}, \eqref{equ07x}, and \eqref{equ07y} represent a nonlinear superposition of two traveling waves
with periodically varying amplitude.
\medskip

$3^\circ$.
The nonlinear Schr\"odinger equation with delay \eqref{eq01} with a more general than \eqref{equ06} potential of the form
\begin{equation}
F(|u|,|\bar u|)=f(z),\quad \ z=g(|u|)-g(|\bar u|),
\label{equ07z}
\end{equation}
where $f(z)$ and $g(w)$ are arbitrary functions,
also admits exact solutions which are determined by formulas \eqref{eq02}, \eqref{equ07} and \eqref{eq02}, \eqref{equ07x}, \eqref{equ07y}.

Setting in \eqref{equ07z} $f(z)=bz$ and $g(w)=w^2$, we obtain the nonlinear Schr\"odinger equation with delay \eqref{eq01} with cubic nonlinearity given by the potential
$$
F(|u|,|\bar u|)=b(|u|^2-|\bar u|^2).
$$

$4^\circ$.
Let us now consider the nonlinear Schr\"odinger equation with delay \eqref{eq01}
and another potential satisfying the condition \eqref{equ06xyz} and having the form
\begin{equation}
F(|u|,|\bar u|)=f(|\bar u|/|u|).
\label{equ08}
\end{equation}

We look for an exact solution of equation \eqref{eq01}, \eqref{equ08} of the form \eqref{eq02}, assuming
\begin{equation}
r=a(x)b(t),
\label{equ09}
\end{equation}
where the functions of different arguments $a=a(x)$ and $b=b(t)$ must be found in the course of further analysis.
Substituting  \eqref{equ09} into system \eqref{eq04}, and then divide
the resulting equations by $r=ab$. Taking into account the type of potential \eqref{equ08}, as a result
we obtain the functional differential equations with delay
\begin{equation}
\begin{aligned}
&-\varphi_t+k\frac{a''_{xx}}a-k\varphi_x^2+f\Bigl(\frac{\bar b}b\Bigr)=0,\quad \ \bar b=b(t-\tau),\\
&\frac{b'_t}b+2k\frac{a'_x}a\varphi_x+k\varphi_{xx}=0,
\end{aligned}
\label{equ10}
\end{equation}
containing functions of different arguments.

The function $\varphi$ is looked for as the sum of two functions
\begin{equation}
\varphi=c(x)+d(t).
\label{equ11}
\end{equation}
Substituting \eqref{equ11} into the system of equations \eqref{equ10} and separating the variables, we come to a mixed
system consisting of one ODE with delay and three ODEs without delay:
\begin{equation}
\begin{aligned}
&k\frac{a''_{xx}}a-k(c'_x)^2=C_1,\\
&d'_t-f\Bigl(\frac{\bar b}b\Bigr)-C_1=0,\\
&2k\frac{a'_x}ac'_x+kc''_{xx}=C_2,\\
&\frac{b'_t}b+C_2=0,
\end{aligned}
\label{equ12}
\end{equation}
where $C_1$ and $C_2$ are arbitrary constants.

Integrating the fourth ODE \eqref{equ12}, and then the second equation, we find functions depending on time
\begin{equation}
b=C_3e^{-C_2t},\quad \ d=[C_1+f(e^{C_2\tau})]t+C_4,
\label{equ13}
\end{equation}
where $C_3$ and $C_4$ are arbitrary constants.
Functions depending on the spatial coordinate are described by the nonlinear system of second-order ODEs
\begin{equation}
\begin{aligned}
&k\frac{a''_{xx}}a-k(c'_x)^2=C_1,\\
&2k\frac{a'_x}ac'_x+kc''_{xx}=C_2.
\end{aligned}
\label{equ14}
\end{equation}

Note that ODEs \eqref{equ14} admit the simple exact solution
\begin{equation*}
a=C_5e^{\beta x},\quad \, c=\gamma x+C_6,
\end{equation*}
where $C_5$ and $C_6$ are arbitrary constants, and $\beta$ and $\gamma$ are the real roots of the algebraic system of equations
$$
k\beta^2-k\gamma^2=C_1,\quad \ 2k\beta\gamma=C_2,
$$
which is reduced to the biquadratic equation.

In the general case, the system \eqref{equ14} using the transformation
$$
\xi=a'_x/a,\quad \ \eta=c'_x
$$
is reduced to an autonomous system of first-order ODEs, which, after eliminating $x$, is reduced to a single first-order ODE (to the Abel equation of the second kind).
\smallskip

$5^\circ$.
The nonlinear Schr\"odinger equation with delay \eqref{eq01} with potential \eqref{equ08}
has also other exact solutions of the form \eqref{eq02} with functions
\begin{equation}
r=C_1\sin(C_2x+\beta_n t+C_3), \quad \ \varphi=A_nx+B_nt+C_4,
\label{equ07xx}
\end{equation}
where $C_1$, $C_2$, $C_3$, and $C_4$ are arbitrary real constants ($C_2\not=0$),
and other parameters are determined by the formulas
\begin{equation}
\beta_n = \frac{2\pi n}\tau, \quad A_n=-\frac{\beta_n}{2kC_2},\quad B_n=-kC_2^2-\frac{\beta_n^2}{4kC_2^2}+f(1),\quad n=0,\,\pm 1,\, \pm 2,\,\dots
\label{equ07yy}
\end{equation}

The solutions described by the formulas \eqref{eq02}, \eqref{equ07xx}, \eqref{equ07yy} represent a nonlinear superposition of two traveling waves, periodic in time and space.
\medskip

$6^\circ$.
The nonlinear Schr\"odinger equation with delay \eqref{eq01} with a more general than \eqref{equ08} potential of the form
\begin{equation*}
F(|u|,|\bar u|)=f(z),\quad \ z=g(|\bar u|)/g(|u|),
\end{equation*}
where $f(z)$ and $g(w)$ are arbitrary constants,
also admits exact solutions, which are determined by the formulas given above in $3^\circ$ and $4^\circ$ of this section.

\subsection{Other nonlinear Schr\"odinger equations with delay. Some remarks}

$1^\circ$.
Using the system of equations \eqref{eq04} it can be shown that the nonlinear Schr\"odinger equation with delay \eqref{eq01} and the potential
\begin{equation}
F(|u|,|\bar u|)=f(|u|^p-c|\bar u|^p),\quad \ c>0,
\label{equ06aa}
\end{equation}
has solution \eqref{eq02} with functions
\begin{equation}
r=C_1\exp(C_2x+\lambda t), \quad \ \varphi=Ax+Bt+C_3,
\label{equ07aa}
\end{equation}
where $C_1$, $C_2$, and $C_3$ are arbitrary real constants ($C_2\not=0$), and other parameters are expressed by the formulas
\begin{equation}
\lambda = \frac{\ln c}{p\tau}, \quad \ A=-\frac{\lambda}{2kC_2},\quad \ B=kC_2^2-\frac{\lambda^2}{4kC_2^2}+f(0).
\label{equ07ab}
\end{equation}

Substituting $f(z)=bz^2$ and $p=1$ into \eqref{equ06aa}, we arrive at the Schr\"odinger equation with delay \eqref{eq01} and cubic nonlinearity determined by a potential of the form
$$
F(|u|,|\bar u|)=b(|u|-c|\bar u|)^2,\quad \, c>0.
$$

Setting $f(z)=bz$ and $p=2$ in \eqref{equ06aa}, we obtain the Schr\"odinger equation with delay \eqref{eq01} and cubic nonlinearity with the potential
$$
F(|u|,|\bar u|)=b(|u|^2-c|\bar u|^2),\quad \, c>0.
$$

$2^\circ$.
The nonlinear Schr\"odinger equation with delay \eqref{eq01} with a more general than \eqref{equ06aa} potential of the form
\begin{equation}
F(|u|,|\bar u|)=f(z),\quad \ z=(|u|^p-c_1|\bar u|^p)g(|u|,|\bar u|),\quad \ c_1>0,
\label{equ06aa*}
\end{equation}
where $f(z)$ and $g(v,w)$ are arbitrary functions,
admits exact solutions, which for $c_1\not=1$ are determined by the formulas \eqref{eq02}, \eqref{equ07aa}--\eqref{equ07ab}, and for $c_1=1$ by the formulas \eqref{eq02}, \eqref{equ07x}--\eqref{equ07y}.

Setting in \eqref{equ06aa*} $f(z)=bz$, $g(v,w)=v-c_2w$, and $p=1$, we obtain the nonlinear Schr\"odinger equation with delay \eqref{eq01} and cubic nonlinearity, given by the potential
\begin{equation}
F(|u|,|\bar u|)=b(|u|-c_1|\bar u|)(|u|-c_2|\bar u|).
\label{equ06xy}
\end{equation}
For $c_1>0$, $c_2>0$, and $c_1\not=c_2$ ($c_1,c_2\not=1$), the nonlinear Schr\"odinger equation with delay \eqref{eq01}, \eqref{equ06xy} admits two exact solutions, which are described by the formulas \eqref{eq02}, \eqref{equ07aa}--\eqref{equ07ab} for $c=c_1$ and $c=c_2$.

Note that any quadratic potential of the general form
$F(|u|,|\bar u|)=b_1|u|^2+b_2|u||\bar u|+b_3|\bar u|^2$, subject to the condition $b_1^2-4b_1b_2\ge 0$, is reduced to a potential of the form \eqref{equ06xy}.
\medskip

\textit{Remark 5.}
A linear Schr\"odinger equation of the special form was studied in \cite{agi2018}, where delay was taken into account in space derivatives.
\medskip

\textit{Remark 6.}
Nonlinear Schr\"odinger equations with distributed delay
containing integral terms were considered in \cite{wu1996,hal1993,zha2011,che2007}.

\section*{Brief conclusion}

In this paper, for the first time, the general nonlinear Schr\"odinger equation is investigated, the potential of which is set using an arbitrary function of two arguments, depending on the unknown function and on the unknown function with delay.
One-dimensional reductions have been described, which lead the equations under consideration to ordinary differential equations and ordinary differential equations
with delay. A number of exact solutions of nonlinear Schr\"odinger equations with delay, which have been expressed in quadratures or elementary
functions, are given. Solutions have been constructed, the amplitude of which varies periodically in time and space.
All obtained results are new. The exact solutions presented in the article can be used to test numerical methods for integrating nonlinear equations of mathematical physics with delay. It is important to note that the exact solutions obtained are valid for an arbitrary function $f(|u|,|\bar u|)$ included in the general nonlinear Schr\"odinger equation with delay \eqref{eq01}, so they can be used for a wide variety of problems by specifying a specific form of this function.







\section*{Declaration of competing interest}

The authors declare that there is no conflict of interest.

\section*{Funding}

This research was supported by the Ministry of Science and Higher Education of the Russian Federation according to the state task project No. 124012500440-9 and No. FSWU-2023-0031.

\renewcommand{\refname}{References}


\begin{thebibliography}{00}
\medskip


\bibitem{Ag1} G.P. Agrawal,  Nonlinear Fiber Optics, 4th ed. Academic Press, New York, 2007.

\bibitem{Ag2} Yu.S. Kivshar, G.P. Agrawal, Optical Solitons: From Fibers to Photonic Crystals, Academic Press, San Diego, 2003.


\bibitem{Ag3} Y. Kodama, A. Hasegawa, Nonlinear pulse propagation in a monomode dielectric guide,
IEEE Journal of Quantum Electronics 23 (5) (1987) 510--524.

\bibitem{Ag4} P.G. Drazin,  R.S. Johnson, Solitons: An Introduction, Cambridge University Press, Cambridge, 1989.

\bibitem{Ag5}  M.J. Ablowitz, P.A. Clarkson, Solitons Nonlinear Evolution Equations and Inverse Scattering, Cambridge University Press, Cambridge, 1991.

\bibitem{Ag6} Yu.S. Kivshar, B.A. Malomed, Dynamics of solitons in nearly integrable systems, Rev. Mod.  Phys. 63 (1989) 763--915.

\bibitem{liu2019}
W.-M. Liu, E. Kengne, Schr\"odinger Equations in Nonlinear Systems, Springer, 2019.

\bibitem{fib2019}
G. Fibich, The Nonlinear Schr\"odinger Equation: Singular Solutions and Optical Collapse, Springer, 2015.

\bibitem{UFN}
S.A. Akhmanov, A.P. Sukhorukov, R.V. Khokhlov,  Self-focusing and diffraction of light in a nonliner medium,
Sov. Phys. Uspekhi 10 (5) (1968) 609--636.

\bibitem{Has1}
A. Hasegawa, F. Tappert, Transmission of stationary nonlinear optical pulses in dispersive dielectric fibers. I. Anomalous dispersion, Appl. Phys. Letters 23 (3) (1973) 142--144.

\bibitem{Has2}
A. Hasegawa, F. Tappert, Transmission of stationary nonlinear optical pulses in dispersive dielectric fibers. II. Normal dispersion. Appl. Phys. Letters, 1973, 23 (4) (1973) 171--172.

\bibitem{Has3}
K. Tai, A. Hasegawa, A. Tomita, Observation of modulational instability in optical fibers,
Phys. Rev. Letters 56 (2) (1986) 135--138.

\bibitem{Pain_1}
P.G. Drazin, S.R. Johnson, Solitons: An Introduction, Cambridge University Press, 1989.

\bibitem{Pain_2}
M.J. Ablowitz, P.A. Clarkson, Solitons Nonlinear Evolution Equations and Inverse Scattering, Cambridge University
press, 1991.

\bibitem{Pain_3}
J. Weiss, M. Tabor, G. Carnevale, The Painleve property for partial differential equations,
J. Math. Phys. 24 (3) (1982) 522--526.

\bibitem{Pain_4}
N.A. Kudryashov, Painlev\'e analysis of the resonant third-order nonlinear Schr\"odinger equation,
Appl. Math. Letters 158 (2024) 109232.

\bibitem{Pain_5}
N.A. Kudryashov, Painlev\'e analysis of the Sasa--Satsuma equation,
Phys. Letters A 525 (2024) 129900.

\bibitem{polzai2012}
A.D. Polyanin, V.F. Zaitsev, Handbook of Nonlinear Partial Differential Equations, 2nd ed., CRC Press, Boca Raton, 2012.

\bibitem{kha2019}
U. Al Khawaja, L. Al Sakkaf,
Handbook of Exact Solutions to the Nonlinear Schr\"odinger Equations, Institute of Physics Publ., Bristol, 2019.

\bibitem{pol2025}
A.D. Polyanin, Handbook of Exact Solutions to Mathematical Equations, CRC Press--Chapman \& Hall, Boca Raton, 2025.

\bibitem{cho2011}
K.W. Chow, T.W. Ng,
Periodic solutions of a derivative nonlinear Schr\"odinger equation: Elliptic integrals of the third kind,
J. Comput. Appl. Math. 13 (2011) 3825--3830.

\bibitem{Kudr_1}
N.A. Kudryashov, A generalized model for description of propagation pulses in optical fiber,
Optik 189 (42) (2019) 52.

\bibitem{Kudr_2}
N.A. Kudryashov, Solitary and periodic waves of the hierarchy for propagation pulse in
  optical fiber, Optik 194 (2019) 163060.

\bibitem{Kudr_3}
N.A. Kudryashov, Mathematical model of propagation pulse in optical fiber with power nonlinearities,
Optik 212 (2020) 164750.

\bibitem{Kudr_4}
N.A. Kudryashov, Solitary waves of the non-local Schr\"odinger equation with arbitrary refractive index,
Optik 231 (2021) 166443.

\bibitem{Kudr_5}
N.A. Kudryashov, Stationary solitons of the generalized nonlinear Schr\"odinger equation with nonlinear dispersion and arbitrary refractive insex,
Appl. Math. Letters 128 (2022) 107888.

\bibitem{Kudr_6}
N.A. Kudryashov, Almost general solution of the reduced higher-order nonlinear Schr\"odinger equation, Optik 230 (2021) 66347.

\bibitem{Bis_1}
Y. Yildirim, Optical solitons to Schrodinger-Hirota equation in DWDM system with modified simple equation integration architecture,
Optik 182 (2019) 694--701.

\bibitem{Bis_2}
E.M.E. Zayed, R.M.A. Shohib, A. Biswas, M. Ekici, A.S. Alshomrani, S. Khan, Q. Zhou, M.R. Belic,
Dispersive solitons in optical fibers and DWDM networks with Schrodinger--Hirota equation, Optik 199 (2019) 163214.

\bibitem{Bis_3}
E.M.E. Zayed, R.M.A. Shohib, M.E.M. Alngar, A. Biswas, L. Moraru, S. Khan, Y. Yildirim, H.M. Alshehri, M.R. Belic, Dispersive optical solitons with Schrodinger--Hirota model having multiplicative white noise via Ito Calculus, Phys. Letters A: General, Atomic \& Solid State Physics 445 (2022) 128268.

\bibitem{Bis_4}
G. Wang, A.H. Kara, A. Biswas, P. Guggilla, A.K. Alzahrani, M.R. Belic, Highly dispersive optical solitons in polarization-preserving fibers with Kerr law nonlinearity by Lie symmetry, Phys. Letters A: General, Atomic \& Solid State Physics 421 (2022) 127768.

\bibitem{Bis_5}
A. Biswas, M.B. Hubert, M. Justin, G. Betchewe, S.Y. Doka, K.T. Crepin, M. Ekici, Q. Zhou, S. Moshokoa, M. Belic,
Chirped dispersive bright and singular optical solitons with Schrodinger--Hirota equation, Optik 168 (2018) 192--195.

\bibitem{Bis_6}
Q. Zhou, M. Xu, Y. Sun, Y. Zhong, M. Mirzazadeh, Generation and transformation of dark solitons, anti-dark solitons and dark double-hump solitons, Nonlinear Dynamics
110 (2) (2022) 1747--1752.

\bibitem{azz2015}
F. Azzouzi, H. Triki, Ph. Grelu,
Dipole soliton solution for the homogeneous high-order nonlinear Schr\"odinger equation with cubic-quintic-septic non-Kerr terms,
Appl. Math. Model. 39 (3--4) (2015) 1300--1307.

\bibitem{wan2020}
X.-B. Wang, B. Han,
Characteristics of rogue waves on a soliton background in the general three-component nonlinear Schr\"odinger equation,
Appl. Math. Model. 88 (2020) 688--700.

\bibitem{lie2020}
X. Liu, H. Zhang, W. Liu,
The dynamic characteristics of pure-quartic solitons and soliton molecules,
Appl. Math. Model. 102 (2022) 305--312.

\bibitem{huy2021}
T.T. Huynh, M. Quan, Q.M. Nguyen,
Fast soliton interactions in cubic-quintic nonlinear media with weak dissipation,
Appl. Math. Model. 97 (2021) 650--665.

\bibitem{wan2023}
H. Wang, Y. Zhang,
Application of Riemann--Hilbert method to an extended coupled nonlinear Schr\"odinger equations,
J. Comput. Appl. Math. 420 (2023) 114812.

\bibitem{mur2025}
M.A.S. Murad, F.M. Omar,
Optical solitons, dynamics of bifurcation, and chaos in the generalized integrable (2+1)-dimensional nonlinear conformable Schr\"odinger equations using a new Kudryashov technique,
J. Comput. Appl. Math. 457 (2025) 116298.

\bibitem{pen2025}
W.-Q. Peng,
Dynamics of rational and semi-rational solutions of the general $N$-component nonlinear Schr\"odinger equations,
Appl. Math. Model. 137 (2025) 115726.

\bibitem{kud2023x}
N.A. Kudryashov, A. Biswas, A.G. Borodina, Y. Yildirim, H.M Alshehri, Painlev\'e analysis and optical solitons for a concatenated model,
Optik 272 (2023) 170255.

\bibitem{arn2023a}
A.H. Arnous, A. Biswas, Y. Yakup, M. Luminita, I. Catalina, G.P. Lucian, A. Asim, Optical solitons and complexitons for the concatenation model in birefringent fibers, Ukr. J. Phys. Optics 24(4) (2023) 4060--4086.

\bibitem{arn2023b}
A.H. Arnous, A. Biswas, A.H. Kara, Y. Yildirim, L. Moraru, C. Iticescu, S. Moldovanu, A.A. Alghamdi, Optical solitons and conservation laws for the concatenation model with spatio-temporal dispersion (internet traffic regulation), J. Eur. Optical Soc. 19 (2023) 2.

\bibitem{polkud2024}
A.D. Polyanin, N.A. Kudryashov, Closed-form solutions of the nonlinear Schrödinger equation with arbitrary dispersion and potential,
Chaos, Solitons \& Fractals (2024) (it will appear).

\bibitem{wu1996}
J. Wu, Theory and Applications of Partial Functional Differential Equations, Springer-Verlag, New York, 1996.

\bibitem{polsorzhu2024}
A.D. Polyanin, V.G. Sorokin, A.I. Zhurov, Delay Ordinary and Partial Differential Equations, CRC Press, Boca Raton--London, 2024.

\bibitem{mel2008}
S.V. Meleshko, S. Moyo,
On the complete group classification of the reaction-diffusion equation with a delay, J.~Math. Anal. Appl. 338 (2008) 448--466.

\bibitem{pol7}%
A.D. Polyanin, A.I. Zhurov, Functional constraints method for constructing exact solutions
to delay reaction-diffusion equations and more complex nonlinear equations,
Commun. Nonlinear Sci. Numer. Simul. 19 (3) (2014) 417--430.

\bibitem{polzhu2014}
A.D. Polyanin, A.I. Zhurov, New generalized and functional
separable solutions to nonlinear delay reaction-diffusion equations, Int. J.
Non-Linear Mech. 59 (2014) 16--22.

\bibitem{polsor2021b}
A.D. Polyanin, V.G. Sorokin, Construction of exact solutions to
nonlinear PDEs with delay using solutions of simpler PDEs without delay,
Commun. Nonlinear Sci. Numer. Simul. 95 (2021) 105634.

\bibitem{aib2021}
M.O. Aibinu, S.C. Thakur, S. Moyo,
Exact solutions of nonlinear delay reaction-diffusion equations with variable coefficients,
Partial Dif. Equ. Appl. Math. 4 (2021) 100170.

\bibitem{polsor2021}
A.D. Polyanin, V.G Sorokin,
Nonlinear pantograph-type diffusion PDEs: Exact solutions and the principle of analogy, Mathematics 9 (5) (2021) 511.

\bibitem{polsor2023}
A.D. Polyanin, V.G Sorokin,
Exact solutions of reaction-diffusion PDEs with anisotropic time delay, Mathematics 11 (14) (2023) 3111.

\bibitem{polzhu2014c}
A.D. Polyanin,  A.I. Zhurov,
Generalized and functional separable solutions to nonlinear delay Klein--Gordon equations, Commun. Nonlinear
Sci. Numer. Simul. 19 (8) (2014) 2676--2689.

\bibitem{long2016}
F.-S. Long, S.V. Meleshko, On the complete group classification of the
one-dimensional nonlinear Klein--Gordon equation with a delay, Math. Meth. Appl. Sci. 39 (12) (2016) 3255--3270.

\bibitem{polsor2022}
A.D. Polyanin, V.G. Sorokin,
Reductions and exact solutions of Lotka--Volterra and more
complex reaction-diffusion systems with delays,
Appl. Math. Letters 125 (2022) 107731.

\bibitem{bro2023}
P. Broadbridge, B.H. Bradshaw-Hajek, A.J. Hutchinson,  Conditionally integrable PDEs, non-classical symmetries and applications,
Proc. R. Soc. A 479 (2023) 20230209.


\bibitem{galsvi2007}
V.A. Galaktionov, S.R. Svirshchevskii,
Exact Solutions and Invariant Subspaces of Nonlinear Partial Differential Equations in Mechanics and Physics,
Chapman \& Hall/CRC Press, Boca Raton, 2007.

\bibitem{polzhu2022}
A.D. Polyanin, A.I. Zhurov, Separation of Variables and Exact Solutions to Nonlinear PDEs, CRC Press, Boca Raton--London, 2022.

\bibitem{agi2018}
D. Agirseven,
On the stability of the Schr\"odinger equation with time delay, Filomat 32 (3) (2018) 759--766.

\bibitem{hal1993}
J.K. Hale, S.M.V. Lunel, Introduction to Functional Differential Equations, Springer, New York, 1993.

\bibitem{zha2011}
Z. Zhao, W. Ge,
Traveling wave solutions for Schr\"odinger equation with distributed delay, Appl. Math. Model. 35 (2011) 675--687.

\bibitem{che2007}
C.-F. Chen, B. Luo,
The freeze of intrapulse Raman scattering effect of ultrashort solitons in optical fiber, Optik 118 (1) (2007) 1--4.

\end{thebibliography}
\end{document}